\documentstyle[twocolumn,prl,aps,epsfig]{revtex}

\newcommand{\be}{\begin{equation}}
\newcommand{\ee}{\end{equation}}
\newcommand{\bea}{\begin{eqnarray}}
\newcommand{\eea}{\end{eqnarray}}
\newcommand{\lp}{\left(}
\newcommand{\rp}{\right)}
\renewcommand{\vec}[1]{{\bf #1}}

\begin{document}
\title{Charge transport and phase transition in exciton rings}
\author{L.V.~Butov$^1$, L.S.~Levitov$^4$, A.V.~Mintsev$^1$,
B.D.~Simons$^5$, A.C.~Gossard$^2$, and D.S.~Chemla$^{1,3}$}
\address{$^1$Materials Sciences Division, E.~O.~Lawrence Berkeley
National Laboratory, Berkeley, California 94720, USA}
\address{$^2$Department of Electrical and Computer Engineering, University
of California, Santa Barbara, CA 93106}
\address{$^3$Department of Physics, University of California at Berkeley,
Berkeley, California 94720}
\address{$^4$Department of Physics,
Center for Materials Sciences \& Engineering, Massachusetts
Institute of Technology, 77 Massachusetts Ave, Cambridge, MA
02139}
\address{$^5$ Cavendish Laboratory, Madingley Road, Cambridge CB3 OHE, U.K.}
\maketitle

\begin{abstract}
The macroscopic exciton rings observed in the photoluminescence
(PL) patterns of excitons in coupled quantum wells (CQWs) are explained by
a series of experiments and a theory based on the idea of carrier
imbalance, transport and recombination. The rings are found to be
a source of cold excitons with temperature close to that of the
lattice.
We explored states of excitons in the ring over a range of
temperatures down to $380$mK. These studies reveal a sharp, albeit
continuous, second order phase transition to a low-temperature
ordered exciton state, characterized by ring fragmentation into a
periodic array of aggregates.
An instability at the onset of degeneracy in the cold exciton
system, due to stimulated exciton formation, is proposed as the
transition mechanism.
\end{abstract}


Bound electron-hole pairs -- excitons -- are light Bose
particles \cite{Keldysh} with a mass comparable or smaller than
that of the free electron. Since the quantum degeneracy
temperature scales inversely with the mass, it is anticipated that
Bose-Einstein condensation (BEC) of excitons \cite{Keldysh} can be
achieved at temperatures as much as six orders of magnitude larger
than the microKelvin temperatures employed in atom condensation
\cite{Cornell,Ketterle}. High quantum degeneracy temperatures and
the possibility to control exciton density by laser
photoexcitation makes the cold exciton gas a paradigm system for
studies of collective states and many-body phenomena. As well as
BEC \cite{Keldysh}, the cold exciton and electron-hole systems are
expected to host a range of novel collective states, such as a
BCS-like condensate \cite{Keldysh2}, paired two-dimensional
Laughlin liquid \cite{Yoshioka}, coupled Wigner solids
\cite{Yoshioka}, and an excitonic charge density wave state
\cite{Chen}.

The theoretical predictions above rely on the experimental
realization of cold exciton gases. Although the semiconductor
crystal lattice can be routinely cooled to temperatures well below
$1$K in He-refrigerators, after decades of effort with various
materials, it appears to be experimentally challenging to lower
the temperature of the exciton gas to even a few Kelvins. The
exciton temperature, $T_X$, determined by the ratio of the energy
relaxation and recombination rates, exceeds by far the lattice
temperature in most semiconductors. For $T_X$ to approach the
lattice temperature, the lifetime of excitons should considerably
exceed their energy relaxation time.

Due to the long lifetime and high cooling rate, the indirect
excitons in CQWs form a unique system in which a cold exciton gas
can be created. The long lifetimes are due to spatial separation
of the electron and hole wells \cite{Lozovik76} while high cooling
rates result from momentum conservation being relaxed in the
direction perpendicular to the QW plane \cite{Ivanov}. The form of
interaction is crucial for the formation of collective exciton
states. In the absence of in-plane forces, the indirect excitons
are dipoles oriented perpendicular to the QW plane. The repulsive
interaction between such dipoles stabilizes the exciton state
against the formation of metallic electron-hole droplets
\cite{Yoshioka,Zhu} and results in screening of in-plane disorder
\cite{Ivanov02}.

The seemingly unique opportunity to realize a cold exciton gas
stimulated great interest and has led to active experimental
studies
\cite{Fukuzawa,Butov94,Butov98,Parlangeli,Butov01,Timofeev,Butov02,Snoke02,SnokeScience}.
Recently, an intriguing observation was made of bright macroscopic
exciton rings in the spatial PL patterns in CQWs
with an onset of fragmentation of the external ring into an
ordered array of aggregates taking place at a few Kelvin
\cite{Butov02}. Luminescence rings were also observed in other CQW
materials \cite{SnokeScience}. Yet neither the rings nor their
fragmentation have been accounted for by theory.

Here, the conundrum of ring formation is resolved by new
experiments, leading to a consistent and compelling picture, based
on the mechanism
of carrier imbalance, transport and recombination.
We conclude that, since the carriers recombining to form the
excitons are in thermal equilibrium, the rings represent a source
of cold excitons.
As such, the rings present a new opportunity to study novel states
of the cold exciton gas.

With this motivation, we have explored states of excitons in the
ring over a range of temperatures down to $380$mK. These studies
reveal a sharp, albeit continuous, second order phase transition
to a low-temperature ordered exciton state. We present a phase
diagram and propose a model for fragmentation involving
hydrodynamical instability due to stimulated exciton formation at
the onset of degeneracy in the exciton system.

The spatial distribution of indirect excitons, observed by
measuring the PL pattern, undergoes a striking transformation as a
function of photoexcitation power, $P_{ex}$. While at low $P_{ex}$
the indirect excitons are observed only within the excitation
spot, at high $P_{ex}$ the most prominent feature in the PL
pattern is a {\bf bright ring} concentric with the laser spot and
separated from it by an annular dark region (Fig.~1).
The radius of the ring, and hence of the dark region, increases
with $P_{ex}$ \cite{Butov02} exceeding 100 $\mu$m at the highest
intensities (Fig.~1a-c).

How can excitons appear so far away from the excitation spot? The
excitons can, in principle, travel in a dark state after having
been excited, until slowed down to a velocity below photon
emission threshold, where they can decay radiatively
\cite{Feldmann}. This mechanism can account for the somewhat more
diffuse inner ring of smaller radius up to tens of microns
observed around the excitation spot \cite{Butov02}. However,
within this framework, a number of qualitative features of the
external ring appear to be difficult to explain. In particular ---
why the observed rings are so sharp, why the ring radius changes
with the gate voltage, and why the rings produced by distant
sources interact before overlapping.

To that end, we are led to assume that the excitons {\it are
generated} within the ring. Indeed a general off-resonance laser
excitation creates charge imbalance in the CQW structure, leading
to carrier transport and recombination at large distances
\cite{Zrenner90}. Most of the hot electrons and holes, created by
the off-resonance excitation \cite{Butov02}, cool down and form
excitons giving rise to PL observed within or near the laser spot.
However, charge neutrality of the carriers excited in CQW is
generally violated mainly due to electrons and holes having
different collection efficiency to the CQW. Overall charge
neutrality in the sample is maintained by opposite charge
accumulating in the doped region outside CQW. We note
parenthetically that the sign of carrier imbalance cannot be
deduced from first principles. We speculate that the current
through CQWs from n-doped GaAs layers (Fig. 1f) creates electron
gas in the CQW, while excess holes are photogenerated in the laser
excitation spot. However, nothing in the discussion and
conclusions below will be affected if electrons are replaced by
holes and vice versa.

The holes created at the excitation spot diffuse out and bind with
electrons, forming indirect excitons. This process depletes
electrons in the vicinity of the laser spot, creating an
essentially electron-free and hole rich region, which allows holes
to travel a relatively large distance without encountering
electrons. At the same time, a spatial nonuniformity of electron
distribution builds up, causing a counterflow of electrons towards
the laser spot. A sharp interface between the hole rich region and
the outer electron rich area forms (Fig.~1e), since a carrier
crossing into a minority region quickly binds with an opposite
carrier to form an exciton. (Recently, possible relation between
charge imbalance of photoexcited carriers and PL ring patterns has
been informally discussed also by I.V. Kukushkin and D. Snoke.)

To summarize our transport model, electrons and holes move in the
CQW plane, each species in its own quantum well, governed by
coupled diffusion equations \be\label{eq:transport} \dot n=D\Delta
n -w n p + J(r) ,\quad \dot p=D'\Delta p -w n p + J'(r) \ee with
$n(r)$, $p(r)$ the electron and hole concentration, $D$, $D'$ the
diffusion constants, and $w$ the rate at which an electron and
hole bind to form an exciton. The source term for holes is
localized at the excitation spot, $J'(r)=P_{ex}\delta(r)$, while
the electron source is spread out over the entire plane,
$J(r)=I(r)-a(r)n(r)$, with $I(r)$ and $a(r)n(r)$ the currents in
and out of CQW. The stationary solution of
Eqs.~(\ref{eq:transport}), displayed in Fig.~1e, indeed shows a
structure of two regions dominated by electrons and holes, and
separated by a sharp interface where exciton density $n_x\propto n
p$ is peaked. The radius of the inner hole region, determined by
the balance between the radial hole flux and electron counterflow
from the periphery, increases with the excitation power, as in the
experiment (Fig.~1a-c). Density variation across the electron-hole
interface can be obtained from a 1D system of stationary
equations, $Dn''=wnp$, $D'p''=wnp$, with the boundary conditions
$Dn_{x\to +\infty}=cx$, $n_{x\to -\infty}=0$, where $c$ is
particle flux normal to the interface and, by symmetry,
$D'p(x)=Dn(-x)$. Interface width
scales as $(DD'/w c)^{1/3}$, making the exciton ring width
practically independent of its radius (cf. Fig.~1a-c).

The exciton ring can be externally controlled by gate voltage:
it expands by a factor of two
after reducing the voltage across the structure from $V_g=1.3$V to
$V_g=1.262$V (Fig.~2h-j).
This is just as one expects since a reduction of transverse
electric field, and hence of the current $I(r)$, depletes
electrons in CQWs, while holes in the inner region remain
practically unaffected.

Experiments with two rings created by spatially separated laser
spots (Fig.~2a-d) reveal the most intriguing prediction of the
transport model: the carrier distribution is strongly perturbed
not only inside, but also outside the ring. As the spots are
brought closer, the rings attract one another, deform, and then
open towards each other (Fig.~2a-c). This happens well before the
rings merge into a common oval-shaped ring (Fig.~2d), suggesting
the existence of ``dark matter'' outside the rings that mediates
the interaction. In the transport model, which readily accounts
for the attraction (Fig.~2e-g), this dark matter is just the
electron flow outside each ring which is perturbed by the presence
of another ring. Electrons in the area between the rings are
depleted more strongly than at the same radial distance in other
directions, which shifts the balance of carrier fluxes to the hole
side. As a result, the electron-hole interface moves further out
between the rings.

The configuration of the electron-hole interface in a stationary
state can be easily obtained by noting that the time-independent
Eqs.(\ref{eq:transport}) yield Poisson equation
$\Delta\Phi=J_{tot}\equiv J'-J$ for the
linear combination of electron and hole densities $\Phi=Dn-D'p$.
The solution
\be\label{eq:Phi=log} \Phi(\vec r)=-\frac1{2\pi}\int \ln |\vec
r-\vec r'|J_{tot}(\vec r')d^2r' \ee describes hole density in the
region $\Phi>0$ and electron density in the region $\Phi<0$. The
interface is thus given by the contour line $\Phi=0$. In
particular, two hole sources of equal strength at $\vec r=\vec
r_{1,2}$ give a family of Bernoulli lemniscata $|\vec r-\vec
r_1|\times |\vec r-\vec r_2|={\rm const}$ (Fig.~2e-g).

By defocussing the source and illuminating a large area (Fig.~3), a
striking new feature emerges which presents a further piece of
compelling evidence for the proposed mechanism. In the indirect
exciton PL pattern, around the localized spots, small rings appear
which mirror the behavior of the outer ring. However, the rings
{\it shrink} at increasing laser power, indicating that
electron-hole contrast here is inverted. This suggests that the
spots reflect pinholes in the CQW barriers, each representing a
localized source of electrons embedded in the hole rich inner
area.
A model with a bright hole source and a weaker electron source
(Eqs.(\ref{eq:transport}),(\ref{eq:Phi=log})), shows the same
qualitative behavior (Fig.~3g). Increasing hole flux (via
$P_{ex}$), as soon as the electron source is enveloped by the hole
rich area, a ring forms around it and then starts to shrink.

The most surprising feature of the spatial PL pattern is {\bf ring
fragmentation} into circular aggregates forming a highly regular,
periodic array over macroscopic lengths, up to ca. $1$mm
(Figs.~1-4).
The aggregates on the ring are clearly distinct from the localized
bright spots generated by the pinholes: In contrast to the latter,
the aggregates move in concert with the ring when the position of
the source is adjusted.
Moreover, in contrast to the spots, the aggregates are {\it cold}:
they do not contain direct excitons (Fig.~4a,b).

The conditions in the ring are optimal for the formation of a very
cold exciton gas. The excitons in the ring are formed by electrons
and holes which travel over macroscopic distances, and thus may
cool to the lattice temperature.
This facilitates characterization of the ordered exciton state
which appears abruptly as the temperature is decreased below ca.
$2$K. The phase diagram, obtained from the azimuthal PL
distribution Fourier spectrum (Fig.~4c), is presented in Fig.~4d.

The ordered low-temperature exciton phase is a new state that has
not been predicted.
The temperature dependence (Fig.~4) indicates that exciton
ordering is a low temperature phenomenon. Interestingly, it is
observed in the same temperature range (below a few Kelvin) as the
bosonic stimulation of exciton scattering \cite{Butov01},
suggesting that the ordering may be an intrinsic property of a
statistically degenerate Bose-gas of excitons. Regarding the
relevance of quantum effects, we also note that
an estimate of the exciton density in the ring based on
the optical shift due to the repulsive dipolar
interaction \cite{Butov01} yields $n \sim 10^{10}$ cm$^{-2}$. This
corresponds to the quantum degeneracy temperature $T_0=(\pi
\hbar^2 n)/(2Mgk_B) \sim 0.2$ K, with $M=0.21\, m_0$ the exciton
mass, and $g=4$ the spin degeneracy of the exciton state in our
CQWs, and to the occupation number of the lowest energy state
$\nu=e^{T_0/T}-1 \approx 0.5$ at $T=0.4$ K \cite{Ivanov}.

At the onset of quantum degeneracy, the rate at which electrons
and holes bind to form excitons will be enhanced due to stimulated
processes: $w\propto 1+\langle \nu\rangle$ in
Eq.(\ref{eq:transport}). To see how this can cause a
hydrodynamical instability, consider a fluctuation in exciton
distribution which leads to growth of exciton density via
stimulated electron-hole binding. The depletion of local carrier
concentration will cause carriers to stream towards the
fluctuation point, attracting more carriers and leading to
subsequent increase in exciton density.

The simplest to analyze is instability of a steady state with
spatially uniform carrier sources and density distribution. To
simplify algebra, we take identical electron and hole parameters
in Eqs.(\ref{eq:transport}), $D'=D$, $J'(r)=J(r)=\alpha(n_0-n)$,
with $n$ the electron (also, hole) density. Adding an equation for
exciton dynamics,
\be\label{eq:x_transport}
\partial_t n_x=D_x\Delta n_x -\gamma(n_x) + w(n_x)n^2
\ee
with $n_x$ the exciton density, $\gamma(n_x)$ the exciton
radiative recombination rate, and linearizing
Eqs.(\ref{eq:transport}),(\ref{eq:x_transport}) over a uniform
steady state, we have
%
$$
\lp\matrix{\delta \dot n-D\Delta \delta n \cr \delta \dot n_{x}\!
-\! D_{x}\Delta \delta n_{x}}\rp = -\lp\matrix{\lambda\! +\! 2w n
& w' n^2 \cr -2w n & \gamma'\! -\! w' n^2}\rp \lp\matrix{\delta n
\cr \delta n_{x}}\rp
$$
%
where $w'=dw/dn_x$, $\gamma'=d\gamma/dn_x$. Stability requires a
positive definite matrix, which gives $\alpha w' n^2 < (\alpha+2w
n )\gamma'$. Since the binding rate $w(n_x)$ diverges near BEC
transition due to the buildup of stimulated processes,
the system becomes unstable at temperatures approaching $T_{BEC}$.
The azimuthal modulation wavelength, selected by the competition
of the instability with particle diffusion, favors the length
scale of the order of the radial exciton distribution width, as
seen in experiment.

Besides the outlined scenario, several other mechanisms might
account for fragmentation. First, we mention a possibility that
exciton dipoles, by tilting spontaneously from the direction
normal to CQWs, can reverse the sign of exciton interaction to
attractive, causing fragmentation \cite{Strecker}. Also, a purely
classical Gann-type effect is possible, in which the radial
current distribution becomes nonuniform and develops a modulation
in the azimuthal direction. However, the perfect periodicity of
the fragments, the phase diagram and, in particular, the low
transition temperature, all suggest a fundamental collective
mechanism most likely of quantum origin.

\vskip5mm \noindent {\bf Acknowledgements}

We thank A.L. Ivanov, L.V. Keldysh, I.V. Kukushkin, D-H Lee, and
P.B. Littlewood for valuable discussions. Communications on the
theory of PL patterns based on exciton transport and recombination
by A.L. Ivanov are greatly appreciated. This work was supported by
the Office of Basic Energy Sciences, U.S. Department of Energy
under Contract No. DE-AC03-76SF00098 and by RFBR. LL
acknowledges support from "Non-Equilibrium Summer Institute at Los
Alamos" and the MRSEC Program of the NSF under Grant No. DMR
98-08941

\begin{figure}
\caption[]{Exciton rings at different photoexcitation
intensity. (a-d) The spatial pattern of the indirect exciton PL
intensity at $T=380$ mK, $V_g=1.24$ V, and $P_{ex}=310$ (a), $560$
(b), and $930$ (c,d) $\mu$W. On (a-c), the area of view is $410
\times 330 \mu$m. (e) Electron, hole, and exciton distribution
predicted by the transport model, Eq.~(\ref{eq:transport}), at two
different excitation intensities. Exciton profile is sharply
peaked at the electron-hole interface. (f) CQW band diagram.}
\label{fig:1}
\end{figure}

\begin{figure}
\caption[flushleft]{External control of exciton rings.
(a-d), Interaction of two exciton rings, the indirect exciton PL
pattern at $T=380$ mK, $V_g=1.24$ V, and $P_{ex}=230 \mu$W per
beam for different separations between the excitation spots. On
(a-d), the area of view is $520 \times 240 \mu$m. (e-g)
Theoretical prediction for the electron-hole interface evolution
at decreasing distance between two point sources. Note ring
attraction in (b,c,f). (h-j), The indirect exciton spatial PL
pattern at $T=380$mK, $P_{ex}=1.4$mW for varying gate voltage:
$V_g=1.3$V (h), $V_g=1.274$V (i), and $V_g=1.262$V (j), with the
area of view $380 \times 320 \mu$m. } \label{fig:2}
\end{figure}

\begin{figure}
\caption[]{Shrinkage of exciton rings to localized bright
spots at nearly homogeneous excitation. (a-f) Spatial PL pattern
for indirect excitons at $T=1.8$ K, $V_g=1.15$ V, and $P_{ex}=160$
(a), 310 (b), 390 (c), 500 (d), 600 (e), and 1130 (f) $\mu$W. On
(a-f), the area of view is $690 \times 590 \mu$m. The excitation
profile along the diagonal for (a-f) is shown in (h). (g)
Theoretical electron-hole interface for a point source of holes
and a weaker point source of electrons. Arrows show interface
displacement at increasing hole number. Note the ring shrinking
around electron source. The shrinkage of ring $A$ (d) of indirect
excitons is detailed in (i) for $T=380$ mK, $V_g=1.155$ V, and
$P_{ex}=77 - 160 \mu$W (from left to right) with defocused
excitation spot maximum moved to a location directly below ring
$A$. On (i-j), the area of view is $67 \times 67 \mu$m for each
image. The shrinkage of the rings is accompanied by the onset of
the direct exciton emission indicating hot cores at the center of
the collapsed rings.} \label{fig:3}
\end{figure}

\begin{figure}
\caption[]{Fragmented ring and localized spots in the
indirect (a) and direct (b) exciton PL image, the area of view is
$410 \times 340 \mu$m. Note that the hot cores with high-energy
direct excitons, present in PL from localized spots, are absent in
the ring fragments. (c) Exciton density Fourier transform peak
height at the fragmentation period vanishes continuously at a
critical temperature. (d) The phase diagram of exciton states in
the external ring. The phase boundary of the ordered phase
observed at the lowest experimental temperatures (solid line)
along with the ring onset region (dashed line) are marked.}
\label{fig:4}
\end{figure}

\end{document}